# Pressure effect on the topologically nontrivial electronic state and transport of lutecium monobismuthide

H. Gu[1,†], F. Tang[1,2,†], Y. -R. Ruan[3,4], J. -M. Zhang[3,4,*], R. -J. Tang[2], W. Zhao[5], R. Zhao[6], L. Zhang[1], Z. -D. Han[1], B. Qian[1,*], X. -F. Jiang[1], and Y. Fang[1,*]

[1]Jiangsu Laboratory of Advanced Functional Materials, Department of Physics, Changshu Institute of Technology, Changshu 215500, China

[2]Jiangsu Key Laboratory of Thin Films, School of Physical Science and Technology, Soochow University, Suzhou 215006, China

[3]Fujian Provincial Key Laboratory of Quantum Manipulation and New Energy Materials, College of Physics and Energy, Fujian Normal University, Fuzhou 350117, China

[4]Fujian Provincial Collaborative Innovation Center for Advanced High-Field Superconducting Materials and Engineering, Fuzhou, 350117, China

[5]ISEM, Innovation Campus, University of Wollongong, Wollongong, New South Wales 2500, Australia

[6]Jiangsu Key Laboratory of Micro and Nano Heat Fluid Flow Technology and Energy Application, College of Mathematics and Physics, Suzhou University of Science and Technology, Suzhou, 215009, China

[†]H. Gu and F. Tang contributed equally to this work.

* Corresponding address: fangyong@cslg.edu.cn, jmzhang@fjnu.edu.cn and njqb@cslg.edu.cn.

Rare-earth monopnictides are predicted to be nontrivial semimetal candidates and show pressure-induced superconductivity. Here, we grow LuBi single crystal and study the magnetization, transport behaviors and electronic band structures to reveal its topological semimetal feature and superconductivity under pressure. At 0 GPa, the quantum oscillations indicate that there are several topologically nontrivial carrier pockets around the Fermi level, among which the hole ones are isotropic in shape, while the electron ones are anisotropic and responsible for the angular magnetoresistance. Upon compression, the superconductivity emerges in the titled compound, showing a similar pressure dependence as that observed in LaBi. Our calculation suggests that the electronic band structures are robust at low- and high-pressure respectively and thus the topological features are always preserved. Besides, the nearly pressure-independent density of state in LuBi indicates that the conventional electron-phonon coupling appears to play a minor role in the superconductivity.

## I. Introduction

Rare-earth monopnictide $R$Pn (Pn=Sb and Bi) that crystallizes in the cubic rock-salt structure, has attracted considerable attention in physics and materials communities, due to its novel physical properties but simple crystal and electronic band structures [1-6]. Thereinto, the most eye-catching hallmarks are extreme magnetoresistance ($XMR$), topologically nontrivial band structure and superconductivity, which not only make $R$Pn a promising candidate for potential applications in the electronic devices, but also pose challenges to understand the fundamental phenomena in nature [1-9].

Large $MR$ of $10^4$~$10^6$% in the $R$Pn series, which is always ascribed to the electron-hole ($e$-$h$) compensation and ultrahigh carrier mobility ($\mu$) [1-6,10-12], has been widely observed at a moderate magnetic field ($B$) and low temperature ($T$). $e$-$h$ compensation and ultrahigh $\mu$ are also crucial for the $XMR$ in a variety of other semimetals, like WTe$_2$, NbAs, MoSi$_2$, $\alpha$-WP$_2$, ZrB$_2$, and so on [13-17]. Here, a fact can't be ignored is that in NbAs the $B$ lifting of topologically protected backscattering, which is proposed as a dominant mechanism for the transport behavior in Dirac semimetal Cd$_3$As$_2$, and Weyl semimetal WTe$_2$ [19,20], also play a part in its magnetotransport [14]. Thus, it can be seen that the two mechanisms aren't mutually-exclusive but sometimes mutually-compatible in certain systems. The latter $MR$ mechanism is suitable for the systems with topologically nontrivial band topology that is always an unevadable topic once the electronic structure of $R$Pn family is referred. Indeed, in early literatures a consensus about the topological and carrier-compensated features of $R$Pn family has reached [1-6], while their contributions to the $XMR$ remains a topic rife with controversies.

Superconductivity is another fascinating issue for $R$Pn series. It was predicted that LaN, ScBi and YBi were low-$T$ superconductors and underwent structural transition under external pressure (*P*) [21,22]. Experimentally, the *P*-induced superconductivity has been observed in LaSb, LaBi and YBi [7-9]. This is reminiscent of the topological superconductors which are of great importance as one source of Majorana fermions for designing new devices in quantum computations [23,24]. Empirically, in light of the energy-momentum dispersion relation, topological superconductors are somewhat similar to topological insulators, both of which share gaps in the bulk but gapless states on the surfaces, suggesting that by tuning some parameters the two exotic states of matter could interconvert [23,24]. In reality, the *P*-induced superconductivity has been reported in several celebrated topological insulators, including $Bi_2Te_3$ family, $\beta$-$Bi_4I_4$, and so on [25,26]. These compounds have been identified as the promising candidates for topological superconductors at one time. While, debates on whether they indeed host topological superconductivity or not has never ceased [25,26]. Nevertheless, discovering exotic compounds with the cross-platform compatibility of topologically nontrivial electronic states and superconductivity remains a promising way to study topological superconductors.

Recently, LuBi was found to be a compensated semimetal with a large *MR* ~ $10^3$%, which shares similarity with LaBi in the electronic band structures [12,27]. Pavlosiuk *et al* verified that there was a clear band inversion with a tiny gap around the X point in the Brillouin zone (BZ), indicating that LuBi could be a topological semimetal [12]. However, Narimani *et al* pointed out that this compound was a topologically trivial metal with no band inversion and zero topological index ($Z_2$ ~ 0), but a nontrivial one under *P* [28]. Thus, it follows that the topological nature of LuBi is full of controversy. Besides, in LaBi, its superconductivity still remains an outstanding issue, namely the theoretical predictions always contradict with the experimental results [7,29]. Since LuBi is an isostructural and isoelectronic analog of LaBi, studying its transport properties and electronic band structures under $P$ could give clues to address these issues.

Here, we report the topological feature and superconductivity under $P$ in LuBi. Nontrivial Berry phase extracted from quantum oscillation suggests that this compound hosts topological electronic state, which is confirmed by the band structure calculations. Our results also reveal that the topological electronic properties do still hold under $P$ no matter whether LuBi crystallizes in the NaCl- or CsCl-type structure. Besides, we find that the $P$-induced superconductivity in LuBi and its $P$-dependence follow LaBi's features. However, the Fermi-level density of state [$N(E_F)$] is almost $P$-independent in NaCl- and CsCl-type structures, which suggests that the conventional electron-phonon coupling could not contribute to the superconductivity in LuBi.

## II. Experimental and Theoretical method

LuBi single crystals were grown by flux method [12]. Resistivity ($\rho$) measurement was performed by a standard four-probe technique. Specific heat ($C_p$) measurement was carried out by the relaxation method. Magnetization ($M$) was measured using a vibrating sample magnetometer. The low $T$ and $B$ were supplied by a Quantum Design physical property measurement system (PPMS-9 T)

High-$P$ transport measurements were performed in a nonmagnetic Be-Cu alloy diamond anvil cell (HMD Corp., Japan) with soft fine NaCl powder acting as the solid transmitting medium. The sample size is about ~ 80×80×10 µm$^3$. $P$ was applied and calibrated with the ruby fluorescence shift at room $T$.

Electronic structure calculations were carried out using density-functional theory (DFT) as implemented in the Vienna *ab initio* simulation package (VASP), where the electron-ion interaction was described by the projector augmented wave (PAW) method [30,31]. For the exchange-correlation part, a generalized gradient approximation (GGA) in the scheme of Perdew-Burke-Ernzerhof (PBE) functional is adopted [32,33]. The kinetic energy cutoff of the plane-wave basis was set to be 250 eV. A 20×20×20 $k$-point mesh was utilized for the BZ sampling for the self-consistent calculations and the Gaussian smearing method with a width of 0.05 eV was used to broaden the Fermi surface. The band structures and Fermi surfaces of LuBi were calculated using a primitive cell.

### III. Results and discussion

#### A. $T$-dependence of zero-field $\rho$ and $C_p$

Figure 1(a) plots the zero-field $\rho$ as a function of $T$ for LuBi, where the current ($I$) flows in the [010] direction and $B$ is applied along the [100] direction. As presented, $\rho$ decreases almost linearly from 300 K down to 50 K, and shows a weak $T$-dependence below 30 K. At 2 K, $\rho$ approximates to 0.07 μΩ·cm, which is comparable to those reported for the Dirac/Weyl semimetals $Cd_3As_2$, TaAs, and so on [18,34]. Besides, the residual $\rho$ ratio (*RRR*) is nearly 136, indicating good metallicity of the grown crystals. The low-$T$ $\rho$ [inset of Fig. 1(a)] can be well fitted by a $\rho=\rho_0+AT^n$ type relation, where $\rho_0$ and A are constants. The obtained $n \sim 2$ differs from those of LaSb ($n \sim 4$) and LaBi ($n \sim 3$) [1,35], generally suggesting that the electronic-correlation scattering dominates the low-$T$ transport properties [13].

Figure 1(b) shows the $C_p$ as a function of $T$. As plotted, it monotonously increases from 2 K and saturates after 200 K, displaying no anomalies hinting to a phase transition. The $C_p(T)$ of LuBi follows Dulong-Petit law at high $T$ and reaches 48.38 J·mol$^{-1}$·K$^{-1}$ at 200 K, which is close to but slightly smaller than the classical limit 6$R$=49.88 J·mol$^-$

$^{-1} \cdot K^{-1}$. Here, $R=8.314$ J·mol$^{-1}$·K$^{-1}$ is the ideal gas constant. The inset of Fig. 1(b) shows the $C_p/T$ versus $T$, which can be well described by $C_p/T=\gamma+\beta T^2+\delta T^4$ [36]. Here, the first and remaining terms are electron and lattice contributions, respectively. The derived Sommerfeld coefficient $\gamma$ comes around 0.3 mJ·mol$^{-1}$·K$^{-2}$. On the other hand, the lattice contribution $\beta$ is estimated to be 1.2 mJ·mol$^{-1}$·K$^{-4}$, yielding the Debye $T$, $\Theta_D = (12\pi^4 NR/5\beta)^{1/3} = 169.2$ K ($N=2$) [36]. Directly fitting the $C_p$ of LuBi to Debye model (not shown), the corresponding $\Theta_D \sim 171.3$ K. The uncertainty places $\Theta_D$ in the vicinity of 170 K, which is smaller than that of Cd$_3$As$_2$ (200 K) but comparable to that of LaBi (165 K). $\Theta_D$ is a measure of the maximum phonon frequency and thus the stability of crystalline lattice, which suggests that LuBi hosts larger compressibility with respect to Cd$_3$As$_2$.

**B. Quantum oscillations and Berry phase**

Figure 2(a) shows the *MR* of LuBi as a function of *B* at different *T*s. The measurement configuration is plotted in the inset of Fig. 2(a), where *B* and *I* are applied along the [100] and [010] directions, respectively. As displayed, the *MR* shows clear Shubnikov-de-Hass (*SdH*) oscillations at $B > 5$ T, revealing high *u* in LuBi. The *SdH* oscillation can be detected even if *T* increases up to 10 K above which it quickly fades away, indicating that the thermal scattering starts to dominate. After subtracting the smooth backgrounds, as shown in Fig. 2(b), the oscillatory *MR* (Δ*MR*) is obtained, which is periodic in 1/*B* due to the energy level quantization [1]. The corresponding fast Fourier transform (*FFT*) spectra is presented in Fig. 2(c), where only three main frequencies (*F*s): $F_\alpha \sim 480$ T, $F_{2\alpha} \sim 960$ T, and $F_\beta \sim 1078$ T are observed. Previously, additional higher *F* components, which aren't revealed in our data possibly due to their weak nature [37], were present in the *FFT* spectra of LuBi [12]. In the inset of Fig. 2(c), the normalized *FFT* amplitudes corresponding to $F_\alpha$ and $F_\beta$ are plotted as a function of $T$, from which the effective electron masses ($m$) are obtained to be 0.18 $m_e$ ($α$ band) and 0.12 $m_e$ ($β$ band) by fitting the thermal damping $R_T$ in Lifshitz-Kosevich (*LK*) formula [1-4,12,37-45]. The obtained $m$s are comparable with those of other *R*Pn reported in early literatures [1-4,12,37-45]. Figure 2 (d) shows the *B*-dependent *M* at different *T*s, where clear De Hass-Van Alphen (*dHvA*) oscillation with its onset *B* as low as 1.5 T is observed at 2 K. Figure 2(e) plots the *M* oscillation components (Δ*M*) as a function of 1/*B*, extracted by subtracting the smooth background as is done for the *SdH* oscillations. An *FFT* [inset of Fig. 2(e)] reveals two oscillation *F*s: 543 and 1200 T which are close to those extracted from the *SdH* oscillation and thus correspond to the $α$ and $β$ pockets, respectively. To further check the topological nature of these two pockets, the corresponding Berry phases (2π$β'$) should be analyzed. Usually, two methods are employed to extract the 2π$β'$ that carries a value of π for relativistic Dirac fermions [15,37]. One is to map the Landau level (*LL*) index fan diagrams, while the other one is to directly fit the *B*-dependent oscillation amplitude with multi-band *LK* formula [15,37]. Here, the 2π$β'$ of LuBi is determined by means of the latter approach. As shown in Fig. 2(f), a function fitting of Δ*M*, where the data can be described as a superposition of damped sin-function for each *F* component, discloses the 2π$β'$ to be 2π(0.50(1)+$δ$) and 2π(0.48(1)+$δ$) for $α$ and $β$ branches, respectively[15,37]. The nonzero 2π$β'$ obtained from *dHvA* oscillation reveals the topologically nontrivial nature of the electronic band structure of LuBi, which is consistent with that extracted from the first-principle calculations shown below.

**C. Angular-dependent *MR* and *FFT* analysis**

To capture the Fermi surface (*FS*) anisotropy of LuBi, we perform the angular ($φ$)-dependent *MR* measurement at 2 K. Figure 3(a)-3(b) depict the simple sketches of the experimental setups, where *I* flows along the [010] axis and *B* rotates in the *ab* (transverse-to-longitudinal configuration, *TLC*) and *ac* planes (transverse configuration, *TC*), respectively. Here, $\varphi$ is measured from the [100] direction. Figure 3(c)-3(d) show the two polar plots of *MR* corresponding to the *TLC* and *TC* measurement modes, where the twofold and fourfold symmetrized *MR* curves are observed. In Fig. 3(c), *MR* is maximal at $\varphi=0°$ (*B*//*a*), decreases continuously as $\varphi$ increases from 0° to 90° due to the decrement of Lorentz force, and then increases again, so that the minima and maxima repeat every 90°. While, in Fig. 3(d), *MR* becomes minimum at $\varphi=0°$, 90°, 180°, and 270°, which coincides well with the fourfold symmetry of the crystal structure of LuBi and reflects a significant FS morphology anisotropy.

Next, we carry out the *MR* measurement at 2 K and different $\varphi$s as a function of *B* to further extract the detail of LuBi′s *FS* topology. Here, the *TLC* measurement mode is employed. As plotted in Fig. 3(e), obvious *SdH* oscillations can be observed under high *B* at all $\varphi$s. After subtracting the nonoscillatory backgrounds, the $\Delta MR$ oscillations [Fig. 3(f)] become much more visible and remain pronounced with $\varphi$ ranging from 0° to 90°, which suggests that the *FS* of LuBi is three-dimensional in essence [46]. The *FFT* spectrum shown in Fig. 3(g) reveals the dominant oscillation *F* as a function of $\varphi$. It is clear that $F_\alpha$ displays a strong $\varphi$ dependence, while $F_\beta$ is almost $\varphi$-independent. Notably, a $F_{\alpha'} \sim 910$ T comes across at $\theta = 30°$, which originates from the contribution of neighboring (010) plane. Due to the cubic crystal symmetry, $F_{\alpha'}$ decreases on further increasing $\varphi$, and finally becomes the fundamental $F \sim 480$ T corresponding to the (010) plane. The same are true for LaBi and PrBi, indicating that they share the similar *FS* topologies [39,46]. Therefore, from the $\varphi$ dependence, there is every reason to believe that $F_\alpha$ is derived from the electron pocket due to its strongly anisotropic feature, and $F_\beta$ corresponds to the extremal area of an almost spherical hole pocket.

Figure 3(h) shows the $\varphi$-dependent $F_\alpha$, where $F_\alpha$ increases with the increasing $\varphi$ up to 45º and then vanishes or merges into another $F$. As reported, in an ellipsoid, its cross-sectional area $A$ and hence the corresponding $F$ as a function of $\varphi$ simply follows $A \sim F \sim \pi ab/[\sin^2\varphi+(a^2/b^2)\cos^2\varphi]^{1/2}$ [4,46]. Here, $a$ and $b$ are the semimajor and semiminor axes of an ellipsoid, respectively. For an ellipsoidal pocket with a large aspect ratio ($a \gg b$), the above formula will be predigested into an inverse cosine [4,46]. As plotted, the $\varphi$-dependent $F_\alpha$ in Fig. 3(h) roughly follows $F_\alpha = F_0/\cos(\varphi-n\pi/2)$ with $n=0$, which suggests that the electron pocket of LuBi is highly elongated.

### E. *P*-induced superconductivity

$\Theta_D$ is a macroscopic reflection of the lattice stability [47]. As referred, LuBi has a $\Theta_D$ comparable to that of LaBi which shows superconductivity under *P*. Besides, being members of *R*Pn family, the two compounds share similarities in band structures [11-12]. Thus, *P*-modification of the electronic band structure and consequently realization of the superconductivity in LuBi come naturally to any of us. Figure 4(a)-4(b) show the *T*-dependent resistance of LuBi under different *P*s from 0.58 to 23.7 GPa (low-*P*, LP) and from 26.2 GPa to 53.0 GPa (high-*P*, HP), respectively. Here, to clearly display the superconducting (SC) phase transition, only the low-*T* resistance is plotted. Figure S1 (see the Supplementary materials [48]) shows the resistance in the *T* interval from 2 to 50 K under different *P*s. In Fig. 4(a), the resistance at 0.58 GPa shows an upturn at low *T*, which is different to that in LaBi but similar to that in Bi ($P > 2.2$ GPa) [49]. Further increasing *P* to 1.8 GPa, the resistance shows an abrupt drop and then quickly decreases to zero around 4.5 K, indicating the emergence of superconductivity in LuBi. The arrow on 1.8 GPa curve defines the SC phase transition *T* ($T_{sc}$) using the *R*=0 criterion [7]. Clearly, $T_{sc}$ firstly increases from 4.5 K (1.8 GPa) to 7.7 K (9.1 GPa) and then decreases to 6.1 K with *P* increasing to 23.7 GPa. Upon further compression, this onset *T*, as shown in Fig. 4(b), is slightly enhanced again to 6.5 K at 26.2 GPa, following by a monotonous decrease at higher $P$.

Figure 4(c) summarizes our experimental results on LuBi in a $T_{sc}$-$P$ phase diagram, where $P$-induced two dome-shaped SC phases can be easily distinguished. Similar case has been observed in LaBi with the two SC states nicely separated by a first-order structural phase transition [7]. For LuBi, as shown in Fig. S2 (see the Supplementary materials [48]), we also find that a structural NaCl- to CsCl-type transition accompanied by a large volume collapse of 6.6% could develop around 27.7 GPa ($P_c$), which agrees well with the early literature [50]. Thus, a LaBi-like $T_{sc}$-$P$ phase diagram emerging in LuBi can be easily understood. Actually, coinciding with the cases in LaSb and LaBi [7,9], $T_{sc}$ in the CsCl-type structure increases a little bit around $P_c$. Despite all this, it's not that structural transition is a fundamental prerequisite for generation of such an exotic state. A typical example is that the superconductivity emerges in NbAs$_2$ and WTe$_2$ involving no structural transition but at the expense of suppressing its *XMR* [51,52]. Figure S3 (see the Supplementary materials [48]) shows the normal-state *MR* of LuBi as a function of *B* under different *P*s, where *MR* is suppressed by approximately three orders of magnitude. Thus, the superconductivity in the NaCl-type LuBi could be somewhat similar to those in WTe$_2$ and NbAs$_2$, since no structural transition has been revealed below $P_c$ as well [50]. Fitting the *B*-dependent *MR* to a power-law relation *MR* = $(\bar{u} \times B)^n$ [51], the *P*-dependent *n* and average carrier mobility $\bar{u}$ are obtained (see the inset of Fig. S3 [48]). As shown, *n* takes the values between 1.95 ($P$=0) and 1.13 ($P$=9.1 GPa) and $\bar{u}$ is theatrically reduced from 2.94 to 0.03 m$^2 \cdot$V$\cdot$s$^{-1}$ by applied *P*, which is the case in NbAs$_2$ [51]. Usually, whenever the *P*-induced superconductivity in Bi-based compounds is referred, a topic cannot be avoided is that whether it is intrinsic, namely the zero resistance is derived from Bi impurity or not [53]. Carefully comparing the $T_{sc}$-

*P* phase diagrams of Bi, LaBi and LuBi, it can be easily found that they indeed share similar SC phase regions with different onset *P*s and $T_{sc}$ [7,49]. Thus, the task before us is to exclude the contribution of Bi impurities to the superconductivity in LuBi. Here, we try to rule out this from six aspects discussed below. 1) no trace of bulk Bi impurities is found from the powder x-ray diffraction pattern and the energy-dispersive x-ray spectroscopy (see Fig. S3 in the Supplementary materials [48]) of the grown LuBi crystals before loading into the high-*P* device. 2) Large *RRR* ~ 136 indicates that the as-grown LuBi is of good quality. 3) Sharp slope of SC transition reveals a bulk effect [53]. 4) The *T*-dependent resistance of Bi crystals shows two jumps around $T_{sc}$ at 3 GPa [49], which is not found in our data. 5) The superconductivity generated from Bi impurities would be readily suppressed by the applied *B* [49], which contradicts with the low-*T* resistance behavior of LuBi [Fig. 4(d)-4(e)] measured under different *B*s at 3 and 11.5 GPa. 6) Fitting the *T*-dependent $H_{c2}$ extracted from Fig. 4(d)-4(e) to the extended Ginzburg-Landau (GL) equation: $H_{c2}(T)=H_{c2}(0)[1-(T/T_c)^2]/[1+(T/T_c)^2]$ [52], as shown in [Fig. 4(f)], the zero-*T* limit of the upper critical field $H_{c2}(0)$ at 3 and 11.5 GPa are determined to be 10.4 and 17.3 T, which are much larger than those of Bi crystal (0.1 T for Bi-II phase and 3.7 T for Bi-III phase) [49]. Here, $T_c$ is determined as the *T*s corresponding to a 90% resistance drop of the normal state [51,52]. All these evidences listed above seemly support the inference that the superconductivity in LuBi is intrinsic.

## F. Electronic band structure under different *P*s

Figure 5 plots the calculated electronic band structures at different *P*s for LuBi, where the red and blue solid circles denote the Lu-*d* and Bi-*p* states, respectively. Figure 5(a)-5(c) display the band structures under 0, 3.0 and 6.6 GPa. Note that the lattice constants employed to perform the band structure calculations can be found in the Supplementary materials [48]. Agreeing with early reports [12], as shown in Fig. 5(a), doubly degenerated band structures with several bands crossing the $E_F$ are observed at ambient $P$, where the conduction and valence ones deeply dip into each other along the high-symmetry Γ-X line, confirming the compensated character of LuBi. Note that an obvious band crossing, at which the Lu-$d$ electrons at very bottom of the conduction bands and the Bi-$p$ states at very top of the valence bands get inverted, is observed around -0.45 eV below $E_F$, indicating that LuBi could be a potential nontrivial semimetal and $B$ lifting of the topologically protected backscattering couldn't take effect on the $MR$ in this compound. Here, we extract the topological invariant $Z_2$ from the parities of the occupied states at all time-reversal-invariant points according to the Fu-Kane criterion [54]. Totally, eight time reversal-invariant points $\Lambda_i=(k_1, k_2, k_3)$ ($i$=1, 2, 3…8 and $k_{1,2,3}$=0 or π) exist in the BZ, by which a nonzero $Z_2$=(1:000) is obtained from the product of all these occupied states, verifying that the electronic structure of LuBi is topologically nontrivial. Further increasing $P$ to 3.0 and 6.6 GPa, the electron band is gradually moved upward and extended to the Γ point, while the two hole bands are hardly shifted. And also, possibly due to the reduced atom distance and then the enhanced atomic interaction, the band inversion becomes much more pronounced. To clearly show the $P$ effect on the electronic band structures, we plot Fermi surface extracted from Fig. 5(a) and Fig. 5(c) in one figure. At 0 GPa, as shown in Fig. 5(d), there are two hole pockets with a nearly spherical $FS$ ($β$ band) and a stretched $FS$ ($γ$ band) in <100> directions centered at the Γ point and three electron pockets with a ellipsoidal $FS$ ($α_1$, $α_2$, and $α_3$ bands) located near the $X$ point of the first $BZ$. When the sample is under 6.6 GPa, only the sizes of the carrier pockets are altered with their shape nearly unchanged. Figure 5(e)-5(g) show the band structures of LuBi with a CsCl-type lattice. Obviously, relative to those with the NaCl-type structure, the shape of these band structures has been significantly changed with the band inversions still existing at the X point, which indicates that the topological nature of LuBi has not been altered by the applied $P$. It is clear that there are one electron band and one hole band in the first *BZ* with the former one crossing the $E_F$ twice and the latter one passing through the $E_F$ three times. Note that only a subtle change can be found in the band structures from Fig. 5(e)-5(g), suggesting that the electronic state is robust in LuBi with the CsCl-type structure. In Fig. 5(h), we only show the *FS* morphology of LuBi at 34.4 GPa, from which one electron pocket and one hole pocket corresponding to the band structures in Fig.5(e) can be observed. It can be seen that the topological nature of LuBi is always kept under *P*, while its electronic band structures are significantly altered after structural phase transition but almost *P*-independent in each crystal phase. Now, one may raise a question: How can the superconductivity observed in LuBi be understood from the band structure calculation? As reported, along with the occurrence of low-*T* superconductivity, the densities of states for WTe$_2$ and NbAs$_2$ steadily increase under enhanced *P* [51-52], which seem to be absent in LuBi [see Fig. 5(a)-5(c)]. Recently, Zhang *et al* also found that the $N(E_F)$ of LaBi couldn't be significantly changed by *P*, indicating that the superconductivity can't be determined by the conventional electron-phonon coupling mechanism but may be derived from either other unknown ways or the residual Bi impurity [29]. These possibilities could be applicable to understand the superconductivity in LuBi as well, since the effect of *P* on the band structure and transport properties of this compound resembles those in LaBi. Besides, even if all the evidences listed above are against the absence of Bi impurity, its contribution to the superconductivity still can't be completely ruled out, due to the resolution limit of instrument on the microscopic composition analysis. Note that the early electronic band structure calculation on LaBi predicts that the superconductivity with its onset *T* less than the observed value can surely emerge in the CsCl-type structure. It thus can be seen that the superconductivity in $R$Pn could indeed be induced by the applied $P$, which is expected to provide a practical material platform for further studying the connection between topological properties and superconductivity.

IV. Conclusion

In summary, we have systematically studied the $C_p$, $\rho$, $M$, $MR$, and electronic band structures of LuBi to reveal its topologically nontrivial electronic properties and $P$-induced superconductivity. Quantum oscillation analysis suggests that LuBi hosts multiple carrier-pockets around the $E_F$ and topologically nontrivial electronic properties. The $\varphi$-dependent $MR$ evidences that the $FS$ of LuBi is 3D and highly anisotropic in essence. Under applied $P$, two SC domes are observed, which resembles those in LaBi. Our calculations show that the band structures of LuBi with NaCl- and CsCl-type structures are quite different, while their topological nature are always kept. Besides, we also found that similar to that in LaBi, the superconductivity in LuBi can't be simply understood by a conventional way. Further endeavor should be activated to reveal the superconductivity in $R$Pn, especially at its high-$P$ phase, since these compounds could be promising candidates for studying the correlation between topologically nontrivial electronic properties and superconductivity.

**Acknowledgments**

This work is supported by the Key University Science Research Project of Jiangsu Province (19KJA530003), National Natural Science Foundation of China (Grant No. 11604027, 11874113 and U1832147), Natural Science Foundation of Fujian Province of China (No. 2020J02018), and Open Fund of Fujian Provincial Key Laboratory of Quantum Manipulation and New Energy Materials (Grant No. QMNEM1903).

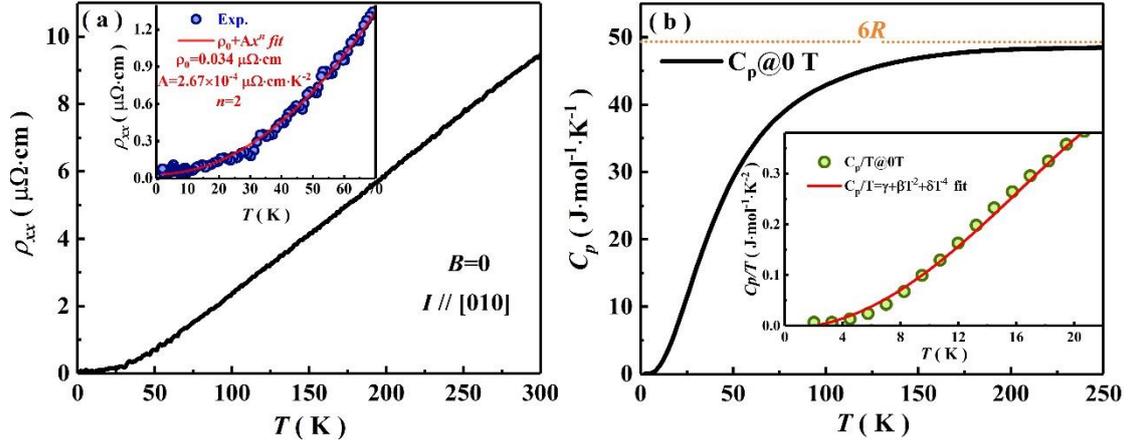

FIG. 1 (a) $T$-dependent $\rho$ at $B=0$. The inset shows the low-$T$ $\rho$ and its fit to $\rho=\rho_0+AT^n$. (b) The $C_p$ as a function of $T$ for LuBi. Inset: $C_p/T$ versus $T$ at low $T$.

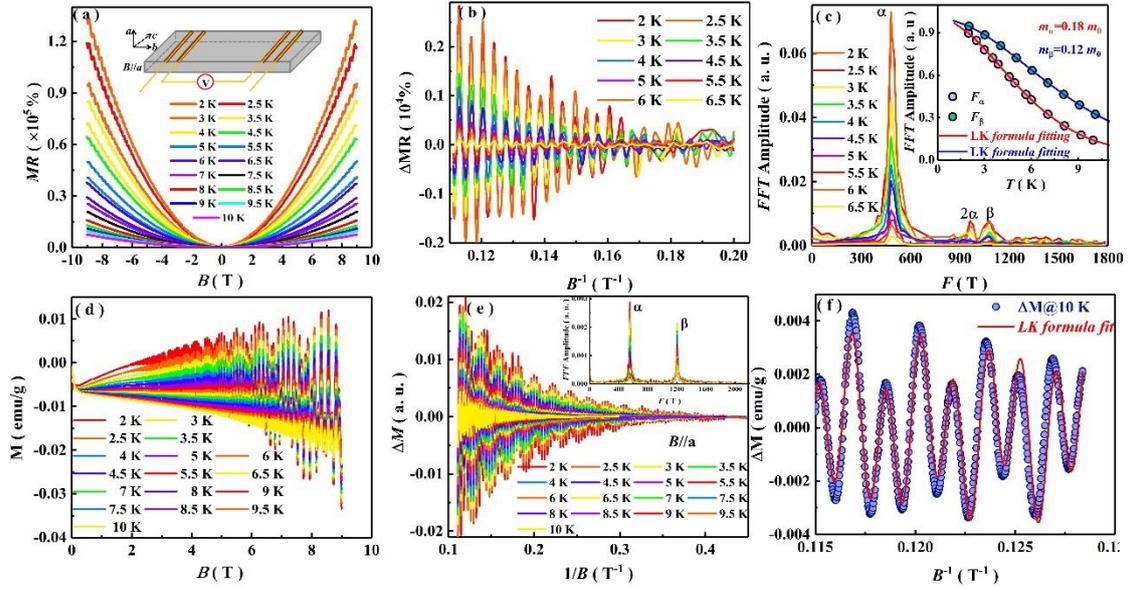

FIG. 2 (a) MR data of LuBi single crystal at various $T$s. Inset: the schematic of the experiment. (b) $\Delta MR$ as a function of $B^{-1}$ at different $T$s. (c) The FFT spectra of SdH oscillations at various $T$s. Inset displays the $T$ dependence of relative FFT amplitudes of the main oscillation $F$. (d) $M$ versus $B$ for $B//a$ at 2-10 K. (e) $B^{-1}$ dependence of dHvA oscillation $\Delta M$. Inset displays FFT spectra of dHvA oscillations. (f) Fitting of dHvA oscillation at 2 K by the multi-band LK formula.

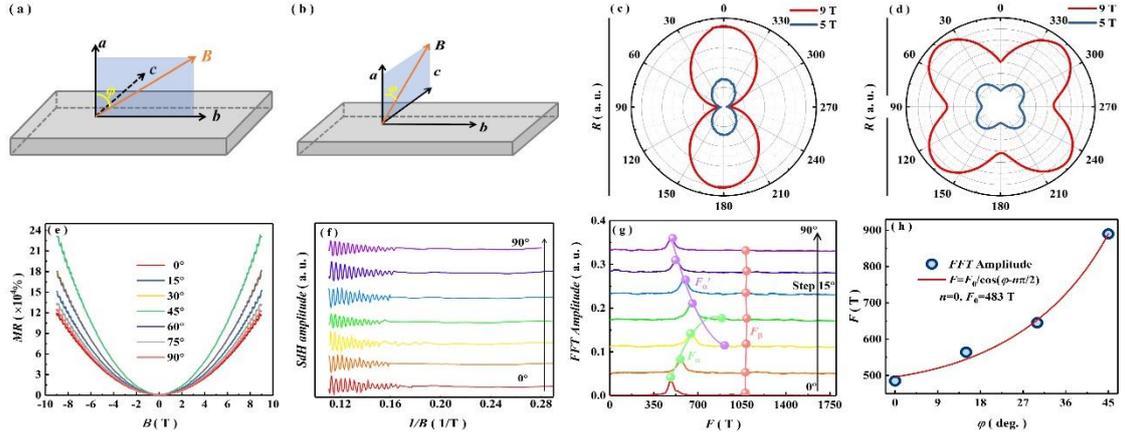

FIG. 3 (a)-(b) show the experimental setups for $\theta$-dependent resistance measurement of LuBi crystal with $I//b$ and $B$ rotating in $ab$ and $ac$ plane respectively. (c)-(d) Polar plot of the resistance at 5 and 9 T in *TLC* and *TC* measurement models respectively. (e) $B$ dependence of *MR* at various $\theta$s. (f)-(g) display SdH oscillation and FFT results at different $\theta$s. (h) $\theta$-dependence of $F_\alpha$. Solid red line are fits to the expression $F=F_0/\cos(\theta-n\pi/2)$, with $F_0=483$ T and $n=0$.

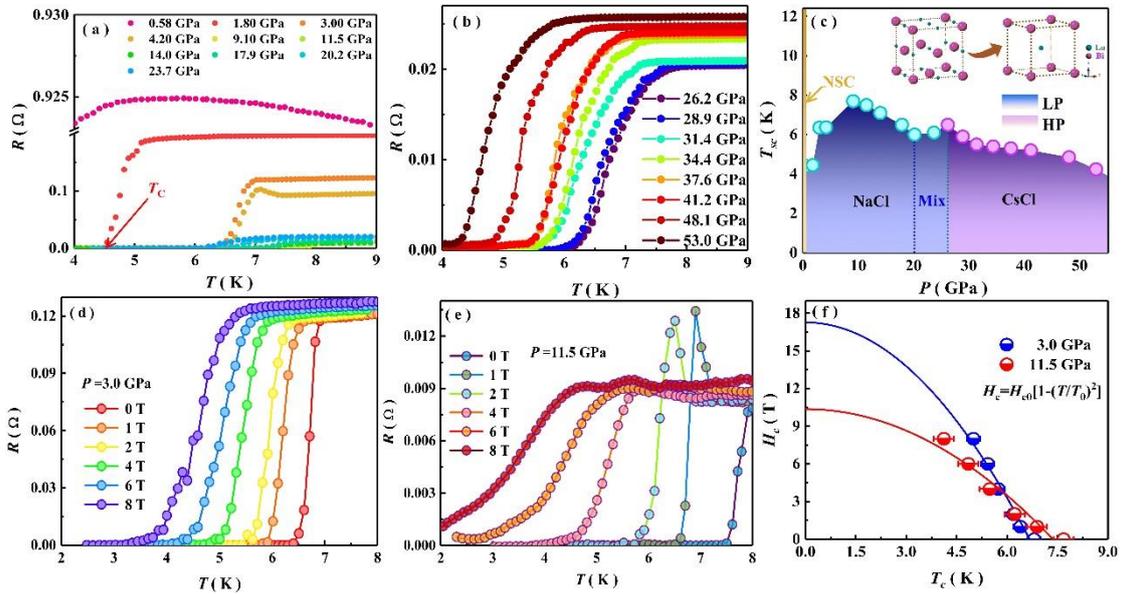

FIG. 4 (a)-(b) $T$ versus low-$T$ resistance at different $P$s. (c) $T_{sc}$-$P$ phase diagram of LuBi. The upper right inset corresponds to the NaCl-type to CsCl-type structural transition. (d)-(e) Low-$T$ resistance at different $B$s, and 3.0 and 11.5 GPa respectively. (f) $H_c$ as a function of $T$ at 3.0 and 11.5 GPa. Solid red and blue lines are fits to GL equation, respectively.

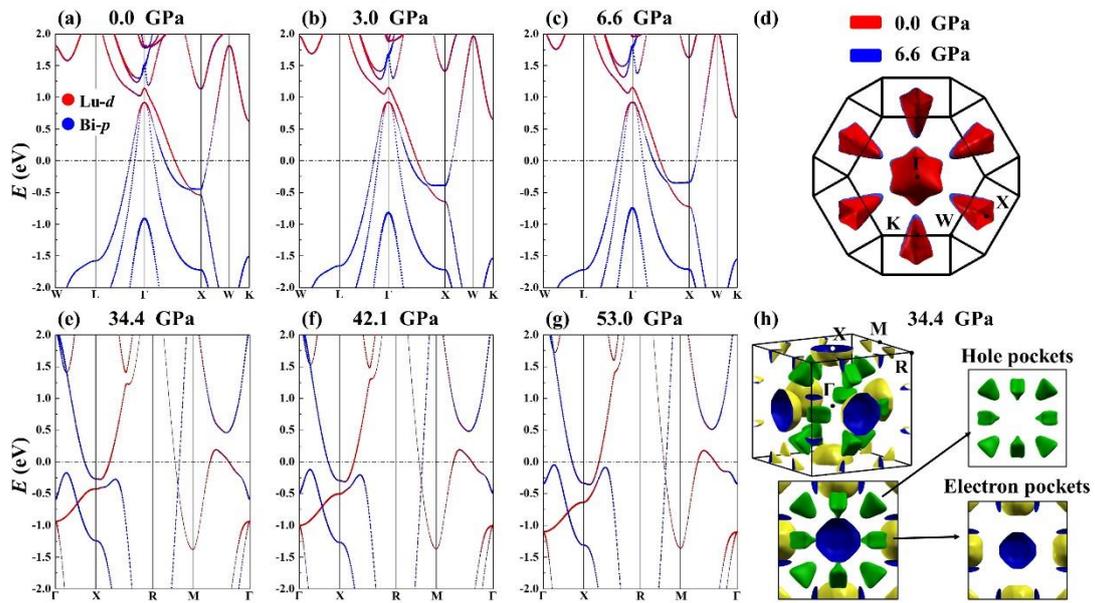

FIG. 5 (a)-(c) Electronic band structures at 0, 3.0 and 6.6 GPa. (d) Fermi surface at 0 and 6.6 GPa. (e)-(g) Electronic band structures at 34.4, 42.1 and 53 GPa. (h) Fermi surface at 34.4 GPa.